\input harvmac.tex

\lref\rWNSP{E. Witten, ``Nonperturbative Superpotentials in String Theory'',
hep-th/9604030.}
\lref\rWI{E. Witten, Nucl. Phys. {\bf B202} (1982) 253.}
\lref\rPKT{P. K. Townsend, Phys. Lett. {\bf B350} (1995) 184.}
\lref\rWSD{E. Witten, Nucl. Phys. {\bf B443} (1995) 85.}
\lref\rAS{A. Strominger, Nucl. Phys. {\bf B451} (1995) 96.}
\lref\rBSV{M. Bershadsky, V. Sadov, and C. Vafa, hep-th/9511222.}
\lref\rVFT{C. Vafa, ``Evidence For F Theory'', hep-th/9602022.}
\lref\rVWOL{C. Vafa and E. Witten, Nucl. Phys. {\bf B447} (1995) 261.}
\lref\rDLM{M. Duff, J. Liu, and R. Minasian,  Nucl. Phys. {\bf B452} (1995)
261.}
\lref\rBB{K. Becker and M. Becker, ``M-Theory on Eight-Manifolds'',
hep-th/9605053.}
\lref\rVDT{C. Vafa, Nucl. Phys. {\bf B273} (1986) 592.}
\lref\rVWDT{C. Vafa and E. Witten, J. Geom. Phys. {\bf 15} (1995) 189.}
\lref\rwitcos{E. Witten, Mod. Phys. Lett. {\bf A10} (1995) 2153.}

\lref\rPKT{P. K. Townsend, Phys. Lett. {\bf B350} (1995) 184.}
\lref\rWSD{E. Witten, Nucl. Phys. {\bf B443} (1995) 85.}

\lref\rJAT{D. P. Jatkar and S. K. Rama, ``F Theory From Dirichlet
Three-Branes'',
hep-th/9606009.}
\lref\rTS{A. A. Tseytlin, ``Self-Duality of Born-Infeld Action and Dirichlet
3-Brane of Type IIB Superstring Theory'', hep-th/9602064. }
\lref\rugw{E. Witten, ``Five-Branes And $M$-Theory On An Orbifold,''
Nucl. Phys. {\bf B463} (1996) 383.}
\lref\rSE{A. Sen, ``Duality and Orbifolds'', hep-th/9604070.}
\lref\rMVFT{D. Morrison and C. Vafa, ``Compactifications of F Theory on
Calabi-Yau Threefolds. I,II'', hep-th/9602114, hep-th/9603161.}
\lref\rbottu{R. Bott and L. W. Tu, {\it Differential Forms In Algebraic
Topology} (Springer-Verlag, 1982).}
\lref\rBSV{M. Bershadsky, V. Sadov and C. Vafa, Nucl. Phys. {\bf B463} (1996)
398.}
\lref\rL{M. Li, Nucl. Phys. {\bf B460} (1996) 351.}
\lref\rMD{M. Douglas, ``Branes within Branes'', hep-th/9512077.}
\lref\rVI{C. Vafa, Nucl. Phys. {\bf B463} (1996) 435.}
\lref\rMH{M. Green, J. Harvey and G. Moore, ``I-Brane Inflow and Anomalous
Couplings on D-branes'', hep-th/9605033.}
\lref\rEZ{E. Zaslow, Comm. Math. Phys. {\bf 156} (1993) 301.}
\lref\rLD{M. Douglas and M. Li, ``D-brane Realization of Super Yang-Mills
Theory in Four-Dimensions'', hep-th/9604041. }
\lref\rSW{N. Seiberg and E. Witten, ``Comments on String Dynamics in
Six-Dimensions'', hep-th/9603003.}
\lref\rCF{P. Candelas and A. Font, ``Duality Between the Webs of Heterotic and
Type II Vacua'', hep-th/9603170.}
\lref\rWP{E. Witten, ``Phase Transitions In M-Theory And F-Theory,''
hep-th/9603150.}


\def\T{{\bf T}}
\def\P{{\bf P}}

\def\O{{\Omega}}

\def\b{{\beta}}
\def\a{{\alpha}}

\def\Ind{{\rm Ind}}

\def\Td{{\rm Td}}
\def\x{{\chi}}

\def\frac#1#2{{#1\over #2}}

%
%
\def\eqnn#1{\xdef #1{(\secsym\the\meqno)}\writedef{#1\leftbracket#1}%
\global\advance\meqno by1\wrlabeL#1}
\def\eqna#1{\xdef #1##1{\hbox{$(\secsym\the\meqno##1)$}}
\writedef{#1\numbersign1\leftbracket#1{\numbersign1}}%
\global\advance\meqno by1\wrlabeL{#1$\{\}$}}
\def\eqn#1#2{\xdef #1{(\secsym\the\meqno)}\writedef{#1\leftbracket#1}%
\global\advance\meqno by1$$#2\eqno#1\eqlabeL#1$$}

\overfullrule=0pt
\Title{hep-th/9606122, HUTP-96/A025,
 IASSNS 96/60}{\vbox{\centerline{Constraints on
Low-Dimensional String Compactifications}}}
\centerline{S. Sethi\footnote{$^\ast$} {Supported in part by the Fannie and
John
Hertz Foundation (sethi@string.harvard.edu).} and C. Vafa }
\medskip\centerline{\it Lyman Laboratory of Physics}
\centerline{\it Harvard University}\centerline{\it Cambridge, MA 02138, USA}
\vskip 0.15in
\centerline{and}
\vskip 0.15in
\centerline{E. Witten\footnote{$^\dagger$}
{Supported
 in part by NSF
Grant PHY95-13835.}}
\medskip\centerline{\it School of Natural Sciences}
\centerline{\it Institute for Advanced Study}\centerline{\it
Olden Lane, Princeton, NJ 08540, USA}

\vskip .3in

We study the restrictions imposed by cancellation of the tadpoles for
two, three, and four-form gauge fields in string theory, M-theory and F-theory
compactified to two, three and four dimensions, respectively. For a large
class of
supersymmetric vacua, turning on a sufficient number of strings, membranes and
three-branes, respectively, can cancel the tadpoles, and preserve
supersymmetry.
However, there are cases where the tadpole cannot be removed in this way,
either because the tadpole is fractional, or because of its sign.
For M-theory and F-theory compactifications, we also explore the relation
of the membranes and three-branes to the nonperturbative space-time
superpotential.

\Date{6/96}
\newsec{Introduction}
String theory compactifications below four dimensions
have not been explored very thoroughly.  This is partly
because compactifications below four dimensions do not at first sight
appear relevant to the observed four-dimensional universe. However, the picture
has changed with recent developments  \rVFT\  that show
that certain four-dimensional
compactifications of F-theory are related to Type IIA and M-theory
compactifications below four dimensions.
Compactifications below four dimensions may also be relevant to
certain four-dimensional questions such as the vanishing
of the cosmological constant \rwitcos.
Finally, given that the string world-sheet is itself two-dimensional,
one naturally expects that some new interesting phenomena
should occur upon compactification of string theory
to two dimensions, or likewise of M-theory to three dimensions,
or of F-theory to four dimensions.

In this paper, we begin to explore some aspects of these compactifications.
In testing one of the predictions of string-string duality,
it was shown in \rVWOL\ that there is a one-loop interaction
in Type IIA superstring theory of the form,
\eqn\efac{\delta S=-\int B\wedge X_8(R),}
where $B$ is the NS-NS two-form of Type IIA which couples to the
string, and  $X_8(R)$ is an
eight-form constructed as a quartic polynomial in the curvature.
In particular, this implies that
compactification on an eight-dimensional manifold $N$ with
\eqn\tadp{I=-\int_N X_8 \not=0,}
results in a one-point  function for the $B$ field of the effective
two-dimensional theory,  destabilizing
the vacuum.  Given the relation between M-theory and Type IIA,  this implies
\refs{\rDLM\rPKT\rWSD}
 that there is a similar term in the effective action
of M-theory where $B$ is replaced by the three-form $C$
which couples to the membrane.  The role of this term in M-theory has been
discussed recently \rBB. Similarly, given the relation
between F-theory compactifications and M-theory \rVFT,
 it follows that there is a similar term in the effective action
of F-theory where $C$ is replaced by $A$, the four-form potential
which couples to the F-theory three-brane.
Note that since three-branes are the only
branes in Type IIB invariant under  $ SL(2,\bf{Z})$, they
play roughly the same role in F-theory that
 membranes play in M-theory, or strings in string theory \refs{\rTS,\rJAT}.

\def\R{{\bf R}}
To cancel the tadpoles present in these compactifications, we include in the
vacuum $I$
strings, membranes, and three-branes in Type IIA, M-theory, and F-theory,
respectively, where $I$ is defined in \tadp.
The world-volumes of these strings, membranes, or three-branes are located
at particular points on $N$, but stretch across the uncompactified space
$\R^n$ (where $n=2,3$, or 4 in the three cases), thus preserving
the $n$-dimensional Poincar\'e invariance.  Some examples of this
kind have been considered recently in \refs{\rugw,\rSE}.
There are two possible obstructions to canceling the tadpole in this way.
One is that $I$ may
be fractional. Since branes only carry integer units of charge, there is
no way to cancel the tadpole by including branes unless $I$ is
an integer.  A more subtle obstruction is that
  $I$ might be integral but negative.
In this case, canceling the tadpole requires anti-branes rather than branes.
This does give a consistent tadpole-free vacuum, but without supersymmetry.
The reason that anti-branes violate supersymmetry is that, first of all,
compactification on the eight-manifold $N$ only preserves supersymmetries
that have a definite eight-dimensional chirality (to be precise,
some but not all
 supersymmetries of the preferred chirality
are preserved in compactification on $N$).
Likewise, inclusion of a brane or anti-brane that stretches across
$\R^n$ at a particular point in $N$ preserves the supersymmetries of
a definite eight-dimensional chirality.  By definition, by ``branes'' we mean
the objects that preserve
 the supersymmetries of the same chirality as those
preserved in compactification on $N$; inclusion of anti-branes
in compactification on $N$ therefore completely breaks the supersymmetry.
To summarize, then, the tadpoles can be canceled in a supersymmetric
fashion by inclusion of branes if and only if $I$ is a positive integer.

For Type IIA compactifications on a Calabi-Yau four-manifold $X$,
 $I$ is proportional to the topological Euler characteristic,
$\x$, of $N$.  To see this, note that in the computation performed in \rVWOL,
one ended up integrating the elliptic genus, defined by taking the odd spin
structure for left-movers and
the sum of all even spin structures for the right-movers,
over the moduli space of elliptic curves.  For compactifications with
space-time
supersymmetry, the sum over all even spin structures gives the same result
as  the odd
spin structure. Thus, the net effect is to get a term proportional
to the path integral with
the odd spin structure for both left- and right-movers, which equals
the world-sheet supersymmetric index $\Tr (-1)^F$  for this theory;
 for geometrical vacua, this is given by the Euler characteristic
$\chi$ of $ N$.  The proportionality constant was fixed in \rVWOL, and
is such that
$$I= {\chi \over 24}.$$

\def\K3{{{\rm K3}}}
 In the following section, we derive some topological constraints on the values
of $\chi/24$ for compactifications on Calabi-Yau four-folds.  We shall also
explore some examples of Type IIA compactifications where $\chi $ is negative.
The case of $ \K3 \times \K3$ is considered as an example of an F-theory
compactification with N=2 supersymmetry requiring twenty-four branes, but also
admitting a dual realization  as an F-theory compactification on
the product of a Calabi-Yau three-fold with a ${\bf T}^2$.
We conclude in section three
 with a discussion of the relation of these
branes to the
non-perturbatively generated space-time superpotential.

\newsec{Topological Constraints}
\subsec{ Calabi-Yau Four-fold Compactifications}

We begin by considering a general Calabi-Yau four-fold, $N$.
On such a space, we can compactify Type IIA, M-theory,
or (if $N$ is elliptically fibered) F-theory to two, three, or four
dimensions.  While the formula $I=\chi/24$ makes it appear
 that the obstruction to cancelling the anomaly
by turning on branes resides generally in ${\bf Z}_{24} $, in actuality we will
show that  the Euler characteristic of a Calabi-Yau four-fold is always
divisible by six, so the obstruction takes values in ${\bf Z}_4$.

\def\ch{{\rm ch}}
To show that the Euler characteristic is always divisible by six, we will use
index theory.  The index of the Dolbeault operator $
{\bar\partial_{E_p}}$ acting on anti-holomorphic forms with values
 in the bundle $E_p$ of holomorphic $ (p,0)$-forms is given by
$$ \Ind( {\bar\partial_{E_p}} ) =  \sum_{q=0}^{n=4}{ (-1)^q \, h^{p,q}, } $$
where $ h^{p,q}$ denotes the dimension of  the cohomology group $ H^{p,q} (N )
$. This index is the twisted arithmetic genus, which can be expressed in terms
of characteristic classes using the Atiyah-Singer index theorem:
\eqn\index{  \Ind ( {\bar \partial_{E_p}} ) = \int_{N} \Td(N) \ch (\O^{p,0}), }
where $ \O^{p,q} $ is the bundle of complex-valued $ (p,q)$-forms on $ N$,
$\Td$ is the Todd class, and $\ch$ is the Chern character.
Since $ c_1=0$, all toplogical indices for the four-fold can be expressed in
terms of the Chern classes $ c_4$ and $ c_2^2$, where $ \int_N c_4=\chi$ is the
Euler characteristic. We begin by considering the untwisted arithmetic genus,
corresponding to $ p=0$ in \index. The Todd genus for a Calabi-Yau four-fold is
given by,
$$  \Td(N)  =   {1\over 720} ( 3 c_2^2 - c_4). $$
This must be two for a Calabi-Yau
four-fold, and therefore,
$$ \int_N c_2^2 = 480 + {\chi\over 3}.$$
  To complete the argument, we only have to consider the case $ p=1$. The
index in this case is given by
$$\Ind ( {\bar \partial_{E_1}} )={1\over 180}\int_N \left(3c_2^2-31 c_4\right)
=8-{\chi \over 6}.$$
The integrality of this index thus implies that $\chi $ is divisible
by $6$.
In the remaining case with $ p=2$, the index,
$$\Ind ( {\bar \partial_{E_2}} )={1\over 120}\int_N \left(3c_2^2+79 c_4\right)
=12+{2 \over 3} \chi ,$$
while further constraining the Hodge numbers, does not provide any new
information about the divisibility of $ \x$.

Can we hope for a
stronger restriction? It is easy to construct an example that shows that the
Euler characteristic of a Calabi-Yau four-fold, while divisible by six,
need not be divisible by 12.
 Consider a Calabi-Yau hypersurface in $ {\bf P}^5$ defined by a
homogeneous polynomial of degree six. A straightforward application of the
adjunction formula gives an Euler characteristic of $2610$ for this space,
and so
$I={2610\over 24}=108  {3\over 4}$; therefore, the obstruction to cancelling
the
tadpole is non-trivial, and takes values
 in $ {\bf Z}_4$. It is easy to construct
other examples of spaces with anomalies that cannot be canceled in a simple
way by turning on branes.

\subsec{The Elliptically-Fibered Case}

A particular class of Calabi-Yau four-folds $N$ is of special interest,
namely, those admitting elliptic fibrations with a section.
Saying that $N$ admits an elliptic fibration means simply
that there is a holomorphic projection
 $ \pi: N \rightarrow B$, whose generic fiber is an elliptic curve;
here $B$ is a complex three-fold, which except in rather special
cases has $c_1$ at least generically positive.

 Such $ N$'s
can be used to compactify F-theory to four dimensions  \rVFT.
Equivalently, we can view such an F-theory compactification
as a Type IIB  compactification on $B$, with
  seven-branes, in which
the modulus of the elliptic fibration describes the variation of the dilaton
and axion fields along $B$.    From that point of view,
the class dual to $12 c_1(B)$ can be identified with
the class where the 7-brane lives \rMVFT.
We can now attempt to rewrite $I=\chi/ 24$ in terms of properties
of $B$ alone.  This turns out to be possible
if the elliptically fibered manifold over $B$ is non-singular.

Before entering into a precise computation, let us note that $I$ will vanish
if $c_1(B)=0$.  Indeed, in that case, $ B$ is a Calabi-Yau three-fold,
and the four-fold $N$ is just   $N=B\times \T^2$, whose Euler characteristic
vanishes.  For this reason, one should expect $\chi(N)$ to be proportional
to $c_1(B)$.  If one assumes that $\chi$ can be written in terms of Chern
classes of $B$, so that
\eqn\reality{I={1\over 24}\chi (N) =\int_B\left( \alpha c_1c_2
+\beta c_1^3\right)}
(we omit a $c_3$ term since it would not vanish when $c_1(B)=0$),
then the coefficients can be determined by considering some simple
examples.
For $N=\K3\times \K3$ we have $\chi(N)/24=24$.  In this case, the base
is $B=\K3\times\P^1$, for which $\int c_1{}^3=0,\int c_1c_2=48$,
so $\alpha ={1\over 2}$.  For another
example, by considering an elliptically fibered Calabi-Yau over ${\bf P}^3$,
for which $\chi/24 =972$, one gets $\beta =15$.
The formula for $I$ is thus
\eqn\eality{I=\int_B\left(15c_1^3+{1\over 2} c_1c_2\right).}

Note further that if the holonomy of $N$ is
$SU(4)$, rather than a subgroup, then by a standard vanishing theorem,
there are no holomorphic differential forms on $N$ except a constant
zero-form and the Calabi-Yau four-form.  There must likewise be
no non-constant holomorphic differentials on $B$ (as a holomorphic differential
on $B$ would pull back to a holomorphic differential on $N$).  The
arithmetic genus of $B$ (that is, $\sum_{i=0}^3(-1)^i{\rm dim}\,H^{i,0}(B)$)
is therefore equal to one.  But the arithmetic genus of a three-fold
is $\int_Bc_1c_2/24$, so for this class of $N$'s we can write
\eqn\eebel{I=12+15\int_Bc_1^3.}
Since the integral in \eebel\ is a positive integer, we see that
for this class of $N$'s, granted that $I$ can be written in terms of Chern
classes, $I$ is always a positive integer, so that supersymmetric
tadpole cancellation is always possible.

\def\co{{\cal O}}
It remains, then, to verify that in fact $I$ can be expressed in terms of
Chern classes; in the process, we will recover the coefficients found
above.  This will be done by a standard computation of
Chern classes. (Much of what is needed below can be found in chapter four
of \rbottu.)
  We recall that the elliptic four-fold $N$ with section
can be described by a Weierstrass equation, which in terms of homogeneous
coordinates $X,Y,Z$ reads $s=0$ where
\eqn\kilk{s=ZY^2-X^3+aXZ^2-bZ^3.}
One can think of $X,Y,$ and $Z$ as homogeneous coordinates on a $\P^2$
bundle $W\to B$.  On the total space of this bundle is a line bundle
that we will call $\co(1)$ (whose sections, when restricted to any $\P^2$
fiber of $W$, are functions of $ X,Y,Z$ homogeneous of degree one).
$Z$ is a section of $\co(1)$, $X$ is a section of $\co(1)\otimes K^{-2}$,
and $Y$ is a section of $\co(1)\otimes K^{-3}$, where $K$ is the canonical
bundle of $B$, pulled back to $N$.  ($a$ and $b$ are sections of $K^{-4}$ and
$K^{-6}$.)  We write the total Chern class of
$B$ as
\eqn\really{C=1+c_1+c_2+c_3}
where in particular $ c_1=-c_1(K)$, and we set $\alpha=c_1(\co(1))$.
The cohomology ring of $W$ is generated,   over the cohomology
ring of $B$, by the element $\alpha$ with the relation
\eqn\kelly{\alpha(\alpha+2c_1)(\alpha+3c_1)=0.}
This relation expresses the fact that the homogeneous coordinates
$Z,X$, and $Y$ have the following two properties: {\it (i)} they
are sections respectively of the line bundles $\co(1)$, $\co(1)\otimes K^{-2}$,
and $\co(1)\otimes K^{-3}$, whose first Chern classes are $\alpha$,
$\alpha+2c_1$, and $\alpha+3c_1$; {\it (ii)} they have no common zeroes,
so that the product of these first Chern classes vanishes.

The relation \kelly\ holds in the cohomology ring of $W$.  Our Calabi-Yau
four-fold $N$ is defined by  the equation $s=0$ where $s$, defined in
\kilk, is a section of $\co(1)^3\otimes K^{-6}$, which is a line bundle
whose first Chern class is $3\alpha+6c_1=3(\alpha+2c_1)$.  Any cohomology
class on $N$ that can be extended over $W$ (and this includes all those
that we will require) can be integrated over $N$
by multiplying it by $3(\alpha+2c_1) $ and then integrating over $ W$.
Since we are in that sense planning to multiply by $\alpha+2c_1$,
the relation \kelly, in the cohomology ring of $N$, can be simplified to
\eqn\jelly{\alpha(\alpha+3c_1)=0.}

Now we wish to compute the Euler characteristic of $N$, or equivalently
its fourth Chern class.  To do this we use the adjunction formula,
which says that the total Chern class of the projective space bundle
$W$ is $C\cdot (1+\alpha)(1+\alpha+2c_1)(1+\alpha+3c_1)$, while the total
Chern class $\overline C$ of $N$ is obtained by dividing this by
$(1+3\alpha+6c_1)$, and is therefore
\eqn\ikko{\overline C=(1+c_1+c_2+c_3){(1+\alpha)(1+\alpha+2c_1)(1+\alpha+3c_1)
     \over (1+3\alpha +6c_1) } .}
The fourth Chern class of $N$, which we will call $\overline c_4$, is simply
the quartic term in the expansion of \ikko.  To compute it, we simply
expand $\overline C$, dropping terms such as $c_1^4$ or $c_1^2c_2$,
which vanish because $B$ is only three-dimensional, and setting
$\alpha^2=-3\alpha c_1$ according to \jelly.
The result is
\eqn\hikko{\overline c_4= 4\alpha c_1(c_2+30c_1^2).}
The Euler characteristic of $N$ is
\eqn\chiis{\chi=\int_N\bar c_4.}
Integration over $N$ can be accomplished by first integrating over the
fibers of $N\to B$, and then integrating over $B$.  In the first step,
one uses the fact that if $F$ is a generic fiber, then $\int_F\alpha=3$.
The ``3'' reflects the fact that the equation $s=0$ is cubic in the
homogenous coordinates $X,Y,Z$, so that the equation $Z=0$ (which is
dual to the cohomology class $\alpha$) defines three points on $F$.  So
we get
\eqn\chias{\chi=12\int_B c_1(c_2+30c_1^2).}
Setting $I=\chi/24$, this is equivalent to the result stated earlier
in \eality.

Before leaving this subject, let us note that the existence of
a formula of this sort reflects the fact that on a D-brane world-volume,
 gravitational and gauge fields induce lower D-brane charges
\refs{\rBSV ,\rMH}.  In particular, it was shown in
\refs{\rBSV,\rMH} that
if a $p$-brane world-volume $K$ has a non-zero first Pontryagin
class $p_1$, there is an induced $p-4$-brane charge given by
${1\over 48}p_1(K)$; if $K$ is a complex manifold
this is the same as $c_2(K)/24$.
In the case at hand,  the seven-brane world-volume
is given by ${\bf R}^4
\times K$ where $K$ (the discriminant of $N\to B$)
is a complex surface  in $B$ whose cohomology class is $12c_1$.  The induced
3-brane charge is thus roughly $\int_B (12 c_1) (c_2(K)/24)$.
One can then, roughly, express $c_2(K)$ as a linear combination of
$c_1(B)^2$ and $c_2(B)$, obtaining for $\chi$ a formula along the lines
above.  The difficulty in making this approach precise is that for {\it any}
elliptic four-fold $N$ of $SU(4)$ holonomy, $B$ has singularities
(corresponding to points at which the fiber of $N\to B$ has a cusp,
that is points at which $a=b=0$ in \kilk), and care is required to determine
the contribution of these singularities to the three-brane charge.
The approach we gave above, in which everything is referred to the $\P^2$
bundle $W$, sidesteps such issues.

The above computation of the Euler characteristic is valid precisely for
elliptic four-folds with {\it (i)} a section, and {\it (ii)} a smooth
Weierstrass model.  If either condition is relaxed, the formula does
not hold.  In particular, to analyze compactifications of $F$-theory
with generic non-trivial unbroken gauge groups, one would like to
understand what happens for elliptic four-folds with a section whose
Weierstrass model is not smooth. We will not try to analyze that case.

\subsec{Examples with $ \chi <0. $}

We have already given examples in which tadpole cancellation fails
because $I$ is not integral.  We here      wish to examine the other
possible obstruction, which is that $I$ might be negative.

Constructing examples of Calabi-Yau four-folds
 with $ \x < 0$ is difficult,
essentially because $ h^{2,1} $, which is the only non-vanishing Hodge number
contributing to $ \x$ with a negative sign, is usually small  in simple
examples.  We wonder whether some of the usual constructions of
Calabi-Yau's as complete intersections in projective spaces or even
in more general toric varieties may always give examples with $\x>0$.

If one is willing to consider more general conformal field theories
such as orbifolds with discrete torsion
(\rVDT; see \rVWDT\ for a recent discussion and geometrical interpretation
of some examples), then it is not difficult, at least in the context
of Type IIA compactification from ten to two dimensions, to find examples
for which the number $I$ of branes needed to cancel the tadpole
is negative.  These are thus superficially reasonable Type IIA
compactifications in which supersymmetry is spoiled by the wrong
sign of the tadpole.  It is plausible that these examples also have
analogs in M-theory and F-theory.

We will consider a simple example for illustration. This example has a
non-zero $h^{2,0}$, and so gives $N=4$ supersymmetry in $d=2$.\foot{
As is customary,
we define the Hodge numbers $h^{p,q}$ of a two-dimensional $N=2$
superconformal field theory in terms of the number of Ramond ground
states with given $U(1)_L\times U(1)_R$ quantum numbers.}

The example is a toroidal orbifold, the target space being
$ \T^8 / {\bf Z}_2 \times  {\bf Z}_2 $.  We describe $\T^8$ via four complex
coordinates $z_i$.   The generator, $ \a$, of the first $ {\bf Z}_2$
acts on the coordinates by, $z_i \rightarrow - z_i$, for $ i=1,2$.
The generator for the second  $ {\bf Z}_2$, denoted $ \b$, acts similarly on
the coordinates $ z_3, z_4$. Without discrete torsion, this model is just
$ \K3 \times \K3$ at a point in its moduli
space with orbifold symmetry.  With discrete torsion, the situation changes.
In this model, the choice of a non-trivial
 discrete torsion is unique. Specifying the phase, $ \epsilon (\a, \b) = -1$,
for a genus one path integral with $\alpha$ and $\beta $ twists,
determines the weights for all other path integral
sectors.  Let us compute the Hodge diamond with this discrete torsion.
The contribution from the untwisted sector is unchanged from the standard
orbifold computation, and is given by,
$$ \left (\matrix{
1 & 0 & 2 & 0 & 1 \cr
0 & 8 & 0 & 8 &0 \cr
2 & 0 & 20 & 0 & 2 \cr
0 & 8 & 0 & 8 &0 \cr
1 & 0 & 2 & 0 & 1 \cr
}\right ),
$$
where the $ (p,q)$ entry is $ h^{p,q}$. The contribution from the three twisted
sectors is more interesting. Let us consider the sector twisted by the element
$ \a\b$ in the quotient group. The fixed point set consists of $ 4^4$ points,
which, without discrete torsion, would contribute $ 4^4$ to $ h^{2,2}$. The
shift of the $ (p,q)$ charges in a twisted sector is described in \rEZ. Because
$ \epsilon (\a, \a \b) = \epsilon (\b, \a \b) = -1$, the fixed points are no
longer invariant under the action of the quotient group, and are projected out;
hence there is no contribution from this sector. The remaining two fixed point
sectors give equal contributions, so we need only consider the sector twisted
by $ \a$. The fixed point set is $ 16$ copies of $ {\bf T}^4$. Because of the
discrete torsion, the only operators which survive projection have (unshifted)
charges given by,
$$\pmatrix{
0 & 32 & 0 \cr
32 & 0 & 32 \cr
0 & 32 & 0 \cr
}.
$$
The total contribution from twisted and untwisted sectors gives a Hodge
diamond,
$$ \left (\matrix{
1 & 0 & 2 & 0 & 1 \cr
0 & 8 & 64 & 8 &0 \cr
2 & 64 & 20 & 64 & 2 \cr
0 & 8 & 64 & 8 &0 \cr
1 & 0 & 2 & 0 & 1 \cr
}\right ).
$$
For this space, $ \x $  is $ - 8 (24) $, so we require $ 8$ anti-branes to
cancel the tadpole. Because $\chi$ is negative, this results in supersymmetry
breaking.

\subsec{F-Theory on $ \K3 \times \K3$}

There are several ways to obtain $N=2$ supersymmetry in four dimensions
from F-theory.
One possibility is F-theory compactification on the product of
a Calabi-Yau
three-fold $C$ with a two-torus.
If $C$ is elliptically fibered over a base $B$,
there are two apparently distinct (but equivalent under further
compactification) versions of F-theory on $C\times \T^2$: one can
consider Type IIB on $C$ itself or on $B\times \T^2$.  We will
be interested in the latter case.
Another $N=2$ model in four dimensions is  F-theory on $\K3\times
\K3$ where one of the $\K3$'s is elliptic
(the latter compactification has been considered
in \rLD ).
  It is natural to ask if the two theories can be dual to one
another.  We will now see that with a proper choice of
three-fold $C$, this is indeed the case.  In the analysis
we will see that in this special case, the branes we have encountered
in this paper are related to more familiar things, and we will get
a check on the proportionality constant in the formula $I=\chi/24$.

One starts from
 the duality of F-theory on $\K3$ to
 heterotic
strings on $\T^2$.  By
compactifying both sides on a $\K3$ down to four dimensions,
we get a duality between F-theory on $\K3\times \K3$ with
heterotic strings on $\K3\times \T^2$.  We consider
the case where we do not excite a gauge bundle over the $ \K3$ on the
heterotic side, but cancel the heterotic world-sheet anomaly
by turning on 24 five-branes (wrapped around $T^2$)
\ref\dmw{M.J. Duff, R. Minasian and E. Witten,
``Evidence for Heterotic/Heterotic Duality'', hep-th/9601036.
}.   The eight dimensional duality
between F-theory and heterotic strings implies that
we need to turn on $24$ three-branes on the F-theory side.
This is, of course, in agreement with the fact that for a
$\K3\times \K3$ compactification, we have $I=\chi/24 =24$.
On the other hand, we can look at $\K3\times \T^2$
compactification of the heterotic string in another way:
we fiber the eight-dimensional duality between the heterotic
string on $\T^2$ and F-theory on K3, to obtain
six-dimensional duality between heterotic strings on $\K3$
and F-theory on a certain  Calabi-Yau three-fold \rMVFT.
This duality can be extended to the case where
the heterotic string has trivial gauge bundle (and 24 fivebranes) by
an appropriate number of blowups on the F-theory side
\rSW.  A description of the
resulting Calabi-Yau three-fold $C$, which has $(h^{11},h^{21})=(43,43)$
has been given in \rCF. We then
compactify this further on a $\T^2$ to four dimensions.
On the heterotic side of course we have the same compactification
as above; we have simply exchanged the 9-10 coordinate with
5-6 coordinate.  On the F-theory side, this geometric symmetry
of the heterotic string
translates to a duality between F-theory on $C\times {\bf T}^2$ -- with this
particular $C$ -- and
F-theory on $\K3\times \K3$.

\newsec{Non-Perturbative Superpotential}

In this paper we have been analyzing compactifications
of F-theory to four dimensions (or
related compactifications of
Type IIA superstring theory or M-theory to  two or three dimensions)
on a Calabi-Yau four-fold $N$.  In the limit that $N$ has large
radius, this gives a four-dimensional vacuum with $N=1$ supersymmetry.
Many of these models, however, generate non-perturbative superpotentials
-- exponentially small in the large radius limit -- because of
instantons that arise from five-brane wrapping \rWNSP.
This occurs when the $F$-theory base $B$ has a divisor $Q$ with
certain properties; it suffices to have $H^{i,0}(Q)=0$, $i=1,2$,
and $H^0(Q,N_Q)=0$, $N_Q$ being the normal bundle to $Q$ in $B$.
We wish in this concluding section to briefly consider a few issues
involving the relation of these non-perturbative superpotentials
to the   branes that have been our main interest in this paper.

First of all, we note a rather surprising fact about the non-perturbative
superpotentials.  Given any $F$-theory base $B$ that does not
have a divisor that can generate a superpotential, we can
 create a complex
manifold $B'$ that does have
such a divisor by blowing up a point  $P\in B$;
we simply let $Q$ be the ``exceptional divisor'' created in the blow-up.
If there is a Calabi-Yau four-fold $N'$ over $B'$ that can be used in
F-theory compactification, then in compactification on $N'$ there
will be a non-perturbative superpotential.

Now, by analogy with what happens for F-theory compactification
to six dimensions \refs{\rMVFT,\rWP}, one might expect that
phase transitions, perhaps of an exotic nature, would occur between
F-theory on $N\to B$ and F-theory on $N'\to B'$.  The superpotential,
however, may be identically zero for F-theory on $N$, while it is
definitely non-zero for F-theory on $N'$.  Although there does not quite
seem to be a contradiction in the fact that there is a superpotential in
$N'$ and none on $N$, we do find it challenging to try to reconcile this
fact with the expectation that a physical transition from compactification
on $N$ to compactification on $N'$ is possible.

We will not meet this challenge here, but point out one fact which
must be part of the story.  The number of branes in compactification on $N'$
is not the same as that in compactification on $N$.  In fact,
if $B'$ is obtained from $B$ by blow-up of a point, then
$\int_{B'}c_1(B')^3-\int_Bc_1(B)^3=-8$.  Thus there are $8\cdot 15=120$
fewer branes if the base is
$B'$ than if it is $B$.  Presumably the
interpretation is that a point $P\in B$ can be blown up only if 120
branes simultaneously arrive at $P$ and annihilate.  There is no
surprise {\it per se} in the fact that some parameters must be adjusted
to make blow-up of a point in the base possible; this is also so
in F-theory compactification to six dimensions \refs{\rMVFT,\rWP}, where
the relevant parameters are complex structure modes of the elliptic
three-fold that is fibered over $B$.  The novelty is merely that
among the parameters that must be so adjusted are the positions of branes.

Another obvious question raised by the presence of branes
in these vacua is how the non-perturbative superpotential depends on the
brane positions.  We note that the non-perturbative superpotential
generated by an instanton associated with a
divisor $Q$ should be expected to get no zero or pole from the motion
of branes in the vacuum
 as long as those remain disjoint from $Q$.  However,
a singular behavior as one of the vacuum branes approaches $Q$ might
be expected, and would be  well worth analyzing.

We would like to thank A. Klemm for valuable discussions.
The research of SS and CV was supported in part by NSF grant PHY-92-18167.
The research of EW was supported in part by NSF grant PHY-9513835.

\bigbreak\bigskip\bigskip

 \listrefs
\end